\documentclass[11pt,twoside]{article}
\usepackage{asp2010}

\resetcounters

\begin{document}

\title{A new sky subtraction technique for low surface brightness data}
\author{Ivan Yu. Katkov$^1$, Igor V. Chilingarian$^{2,1}$
\affil{$^1$Sternberg Astronomical Institute, Moscow State University, 13 Universitetski prospect, 119992, Moscow, Russia}
\affil{$^2$Centre de Donn\'{e}es astronomiques de Strasbourg -- Observatoire de Strasbourg, CNRS UMR 7550, Universit\'{e} de Strasbourg, 11 Rue de l'Universit\'{e}, 67000 Strasbourg, France}
}

\begin{abstract}
We present a new approach to the sky subtraction for long-slit spectra
suitable for low-surface brightness objects based on the controlled
reconstruction of the night sky spectrum in the Fourier space using twilight
or arc-line frames as references. It can be easily adopted for
FLAMINGOS-type multi-slit data. Compared to existing sky subtraction
algorithms, our technique is taking into account variations of the spectral
line spread along the slit thus qualitatively improving the sky subtraction
quality for extended targets. As an example, we show how the stellar metallicity and
stellar velocity dispersion profiles in the outer disc of the spiral galaxy
NGC5440 are affected by the sky subtraction quality. Our technique is used
in the survey of early-type galaxies carried out at the Russian 6-m
telescope, and it strongly increases the scientific potential of large
amounts of long-slit data for nearby galaxies available in major data
archives.
\end{abstract}

\section{Introduction}

Low-surface brightness ($\mu_B>23$ mag/arcsec$^2$) outer regions of galaxies
contain crucially important information for understanding the properties of
their extended discs and dark matter haloes. Brightness profiles of dwarf
early-type galaxies whose mean surface brightness is correlated with the
luminosity, can be entirely in the low-surface brightness regime. Analysis
of absorption line spectra at such surface brightness levels is often
hampered by systematic errors of the sky subtraction that sometimes may lead
to wrong astrophysical conclusions. Therefore, in order to analyse deep
spectral data, it is important to improve the sky subtraction technique.

Here we present a new approach to the sky subtraction for long-slit spectra
based on the controlled reconstruction of the night sky spectrum in the
Fourier space using twilight or arc line frames as references.

\articlefigurethree{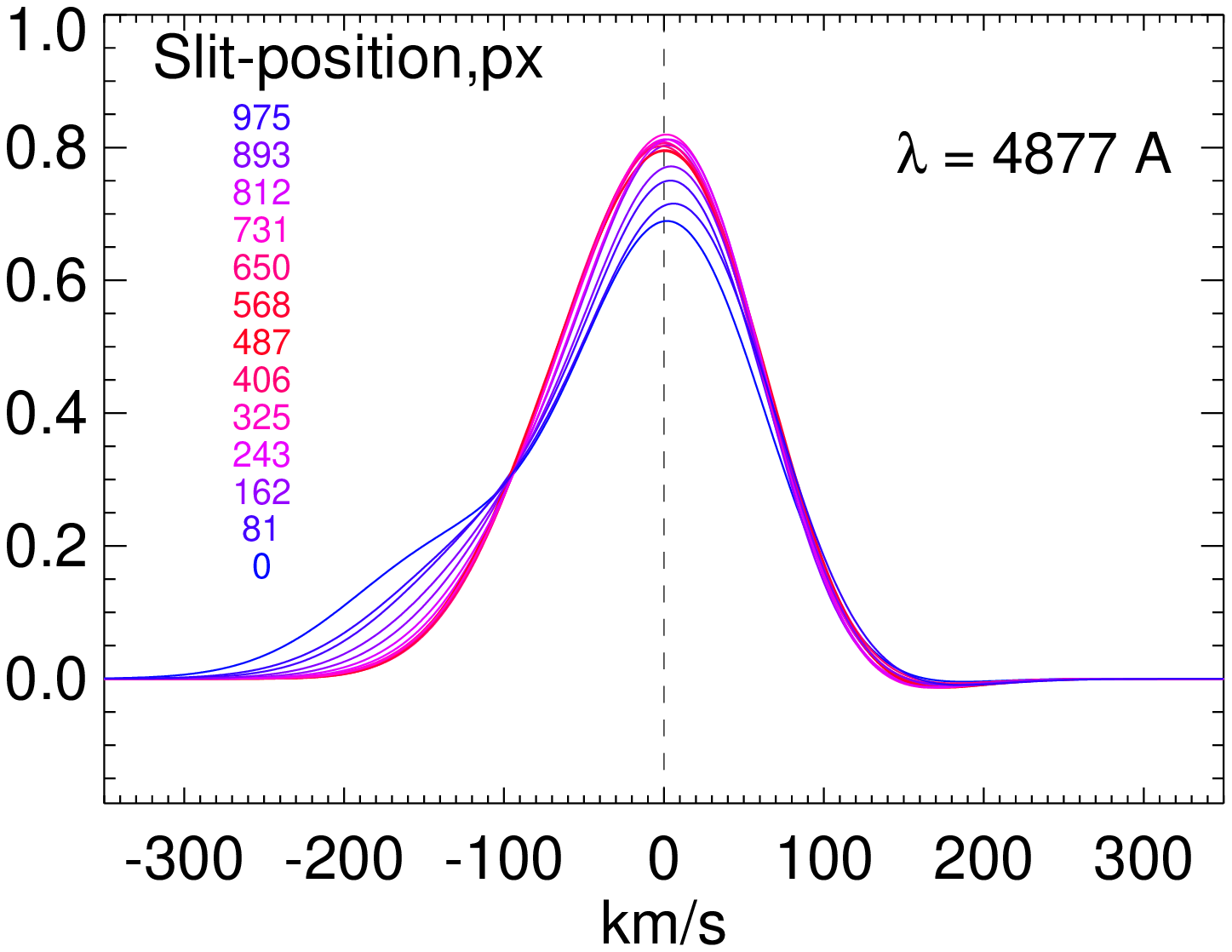}{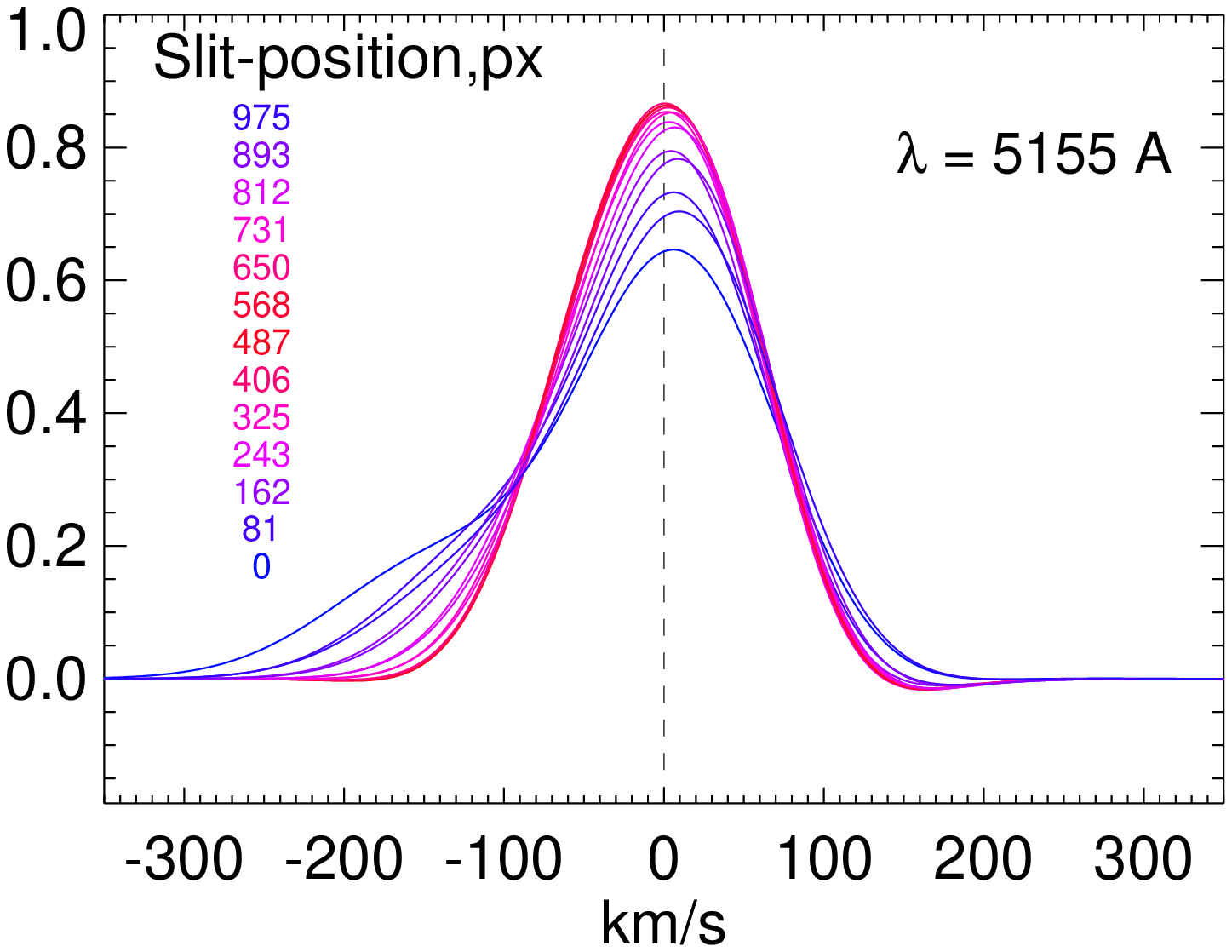}{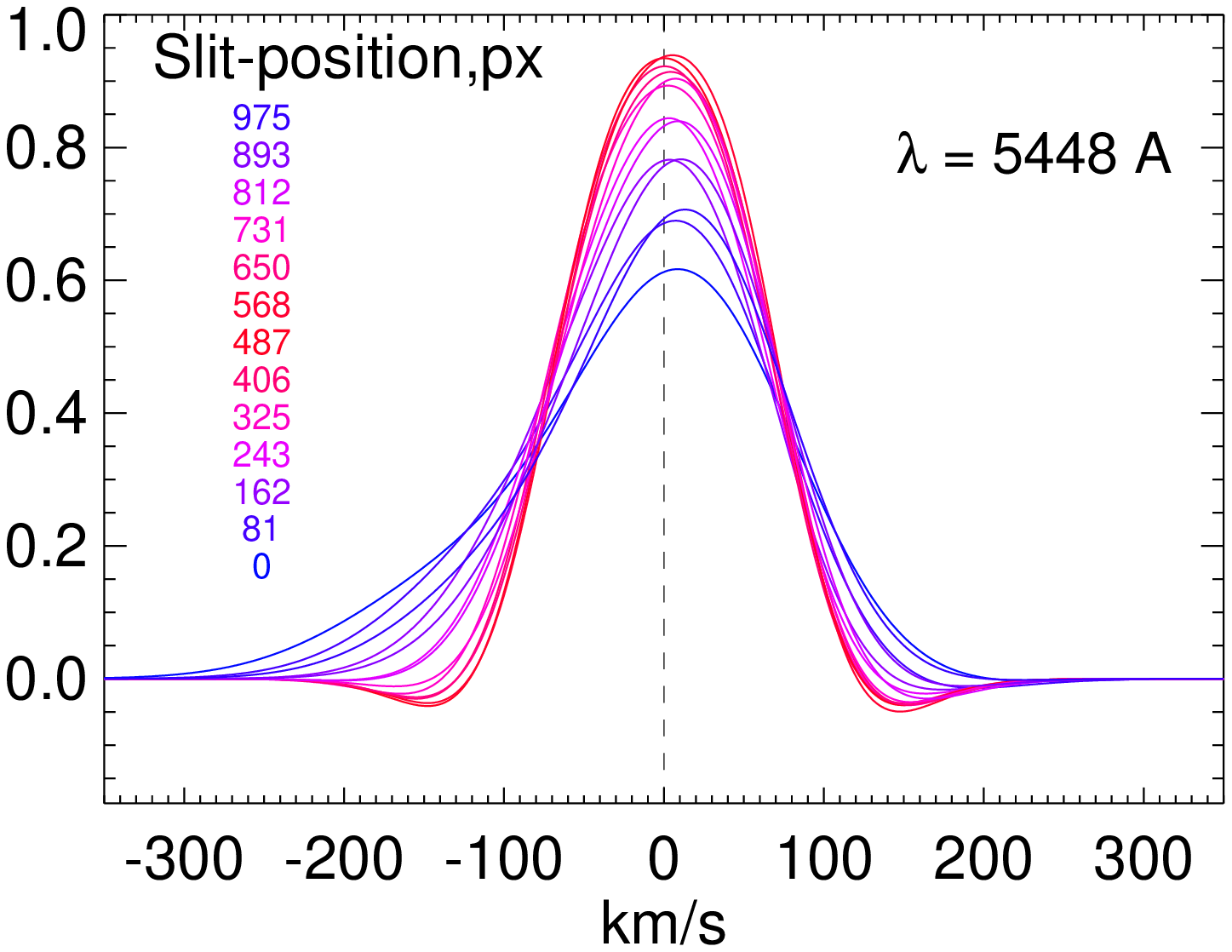}
{fig1_lsf_shape}{An example of the LSF shape of the SCORPIO reconstructed
from the twilight frame at different wavelength and slit positions. We used
the Gauss-Hermite LSF representation. One can see that the profile asymmetry
increases towards the outer slit regions. There is also a notable change of
the overall spectral resolution from blue to red.}

Due to optical distortions, the shape of the spectral line spread function
(LSF) in a long-slit spectrograph varies along the wavelength range as well
as along the slit. In Fig.~\ref{fig1_lsf_shape}, we provide an example of
the LSF shape of the SCORPIO \citep{AM05} universal spectrograph at the
Russian 6-m telescope reconstructed from the twilight frame (i.e. the Solar
spectrum). The LSF is slightly asymmetrical and cannot be described by the
Gaussian function, a usual parametrization in most data reduction packages.
Here we use the Gauss-Hermite representation \citep{vdMF93} up-to the 4th
order moment that allows one to describe first-order differences of the line
profile from the Gaussian shape. These LSF variations affect the night sky
spectrum which is subtracted from science frames during the data reduction.
On the Fig.~\ref{fig2_objspec} we show a reduced long-slit spectrum of the
spiral galaxy NGC 5440 before the sky subtraction step.

\articlefigure{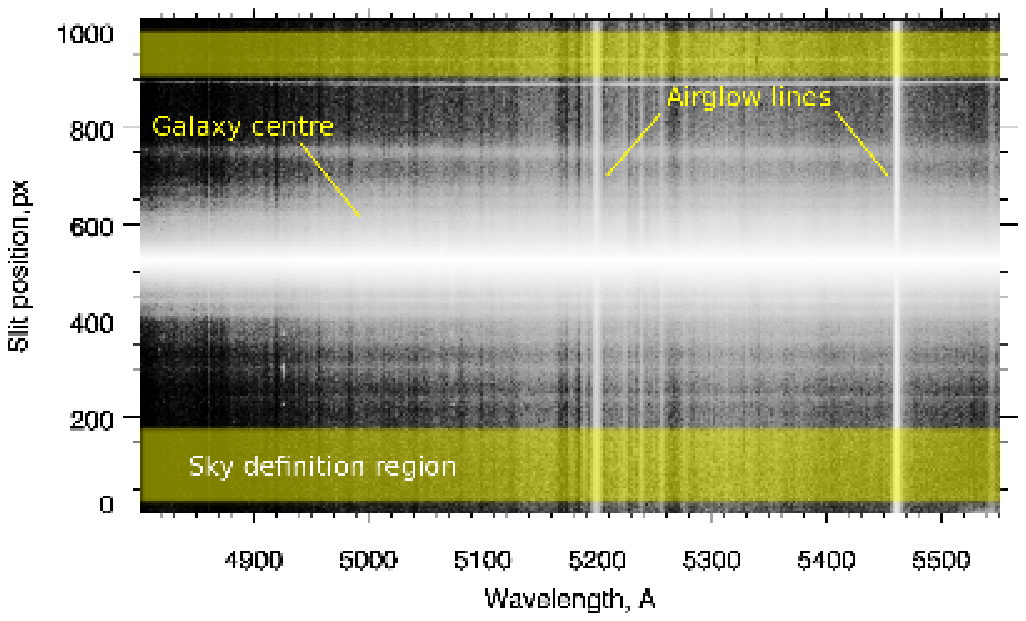}{fig2_objspec}{A
reduced long-slit spectrum of the spiral galaxy NGC~5440 before the sky
subtraction step. Yellow areas denote a region of the frame used to
construct the night sky spectrum used for the sky subtraction. Taking into
account the profile variation shown in Fig.~\ref{fig1_lsf_shape}, it is
clear that the intrinsic LSF shape in these region will differ from that in
regions of the galaxy closed to the slit centre.}

\section{The sky subtraction algorithm}

In the traditional sky subtraction technique implemented in most standard
data reduction packages (IRAF, MIDAS), the night sky spectrum is constructed
from the outer regions of the slit which are (supposedly) free of the galaxy
light as a $\kappa$-sigma clipped average. Then it is subtracted at every slit
position.

The new algorithm presented here is an improvement of a method proposed in
\citet{Chilingarian+09b} aimed at increasing its stability with certain
features taken from the technique by \citet{Kelson03}. However, compared to
the latter method, our approach allows one to take into account empirically
the variations of the LSF along the slit.

The new technique includes several steps:
\begin{enumerate}
 \item An oversampled sky spectrum is created from the non-linearized spectra using the wavelength
solutions in order to perform the pixel-to-wavelength coordinate mapping. Then it is approximated
using a b-spline. This approach was proposed by \citet{Kelson03} to improve the sky subtraction
in undersampled datasets.
  \item At every position along the slit, we change the LSF shape inside this night sky spectrum using a Fourier-based technique into the LSF at that slit position. The observed sky
spectrum is a convolution of a true spectrum with the LSF:
  \begin{equation}
  \begin{array}{lr}
      R(\lambda,y)=R_0(\lambda)*LSF(\lambda,y); & S(\lambda,y)=S_0(\lambda)*LSF(\lambda,y),
  \end{array}
  \label{eq_r_s_spectra} 
  \end{equation}
where $R(\lambda,y)$ is a template spectrum (high signal-to-noise
twilight frame), $S(\lambda)$ -- the night sky spectrum. Then according to the convolution
theorem the ratio between the Fourier transforms of the template spectrum and the object spectrum is a constant function on position along slit $y$:

\begin{equation}
      \frac{FFT(S(\lambda,y))}{FFT(R(\lambda,y))} = \frac{FFT(S_0(\lambda))}{FFT(R_0(\lambda))} = \frac{FFT(S(\widetilde{y},\lambda))}{FFT(R(\widetilde{y},\lambda))} = F(\lambda),      
  \label{eq_fft}
\end{equation}
where $\widetilde{y}$ -- position at the sky definition region.
The night sky spectrum at current position along slit can be expressed as follows:  
  \begin{equation}
      S(y,\lambda) = FFT^{-1} \left( \frac{FFT(S(\widetilde{y},\lambda))}{FFT(R(\widetilde{y},\lambda))} \cdot FFT(R(y,\lambda))  \right).
  \label{eq_skypos}
  \end{equation} 

  \item The $b$-spline parametrization provides the necessary regularisation for the numerical stability of this procedure.
\end{enumerate}

\section{Usage example and perspectives of the method}

In Fig.~\ref{fig3_comparison} we present the result of the data analysis of
a long-slit spectrum of NGC~5440 for the two sky subtraction techniques. We
fitted the reduced sky subtracted spectra with high resolution stellar
population models with the {\sc nbursts} full spectral fitting technique
\citep{CPSA07,CPSK07} and extracted kinematical (radial velocity and
velocity dispersion) and stellar population (age and metallicity) parameters
along the slit. The radial profiles of velocity dispersion and metallicity
are shown in Fig.~\ref{fig3_comparison}. While the measurements are very
similar near the galaxy centre, they differ notably in the peripheral
regions. With the new sky subtraction technique the uncertainties are lower,
and the general trend of the galaxy metallicity gradient correspond to the
physical expectations. The traditional sky subtraction technique possesses
systematic errors in the low surface brightness regime, which propagate to
the data analysis and may result in misleading astrophysical conclusions.

\articlefiguretwo{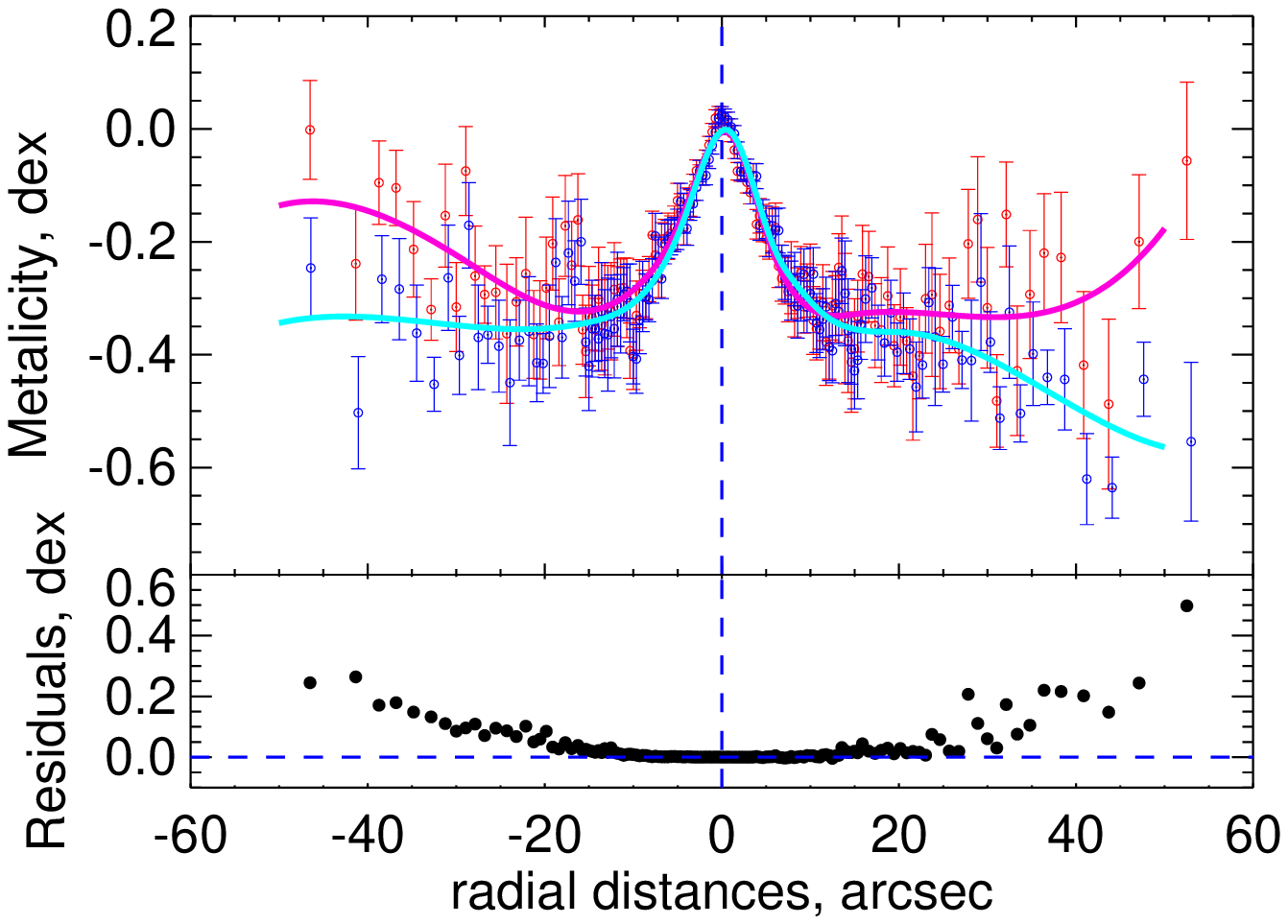}
{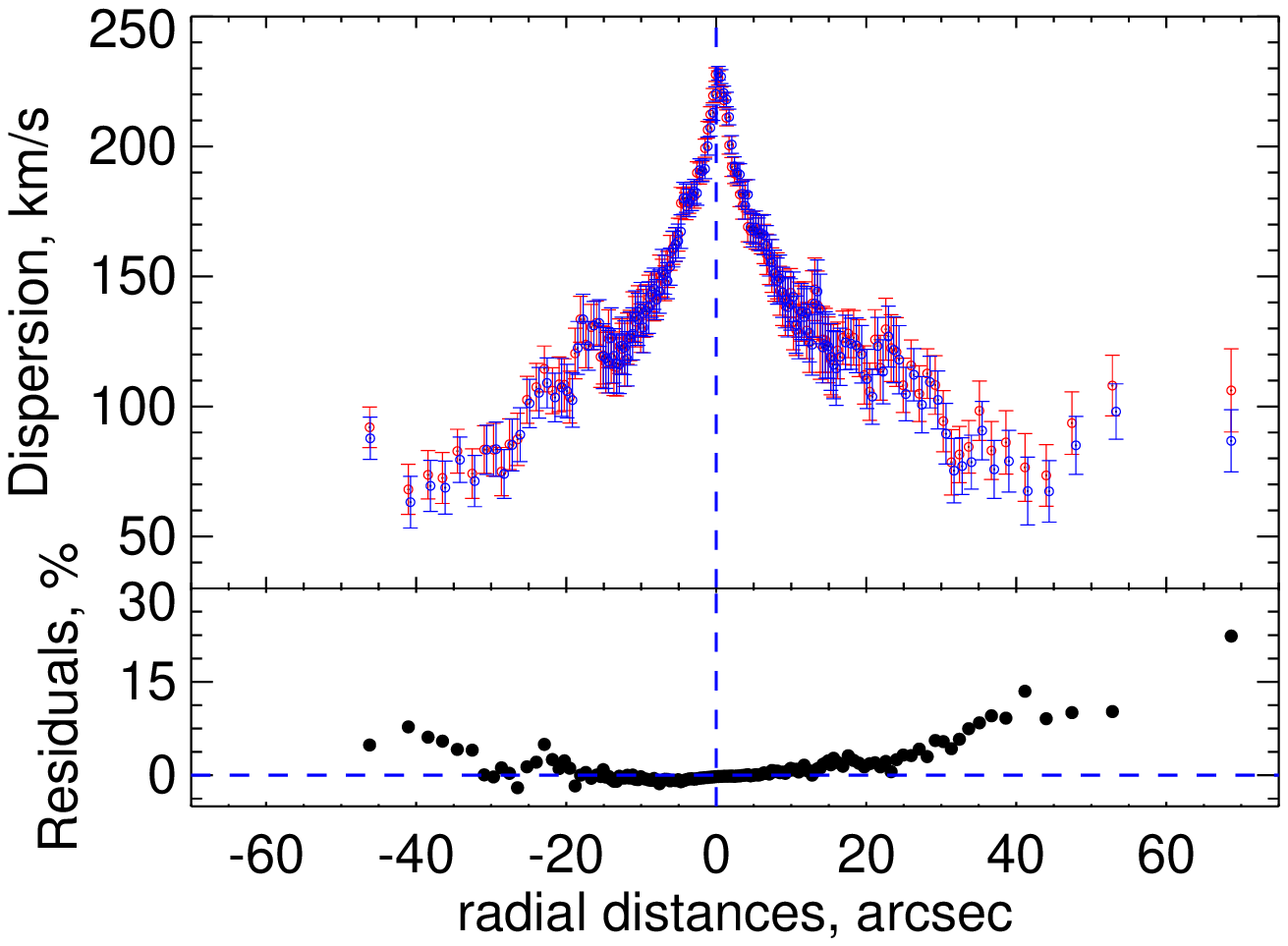}{fig3_comparison}{Comparison between
traditional technique and our deconvolution. The two left panels display the
internal velocity dispersion profiles of NGC~5440. Blue data points are for
the new technique, while red ones are for the ``classical approach''.
Differences between the two approaches are displayed in the bottom panel.
The two right panels display the stellar metallicity profiles of NGC~5440
using the same symbols and colours as on the left panels.}

Our sky subtraction technique is adopted in the survey of nearby lenticular
galaxies (P.I.: prof.~Zasov, Moscow State University) carried out at the
Russian 6-m telescope using the SCOPRIO spectrograph. Our approach can be
easily modified for multi-slit spectroscopic data with parallel slits
(``FLAMINGOS''-type spectra). 

Since out technique improves the quality of data analysis at low
signal-to-noise ratios, it can also be used to re-reduce and re-analyse
long-slit spectroscopic datasets for hundreds of galaxies obtained with
different telescopes and publicly available in data archives.

\acknowledgements IK thanks the ADASS organizing committee for the provided
financial aid and RFBR for covering the remaining travel expenses.

\bibliographystyle{asp2010}

\bibliography{P110}

\end{document}